# Development and characterization of $YBa_2Cu_3O_{7-x}/SrTiO_3$ nanomembrane platform for silicon photonics integration

J.S. Madhira[1,2], G. Potemkin[3], B. Kardynal[1,2], D. Grützmacher[1,2], T. Schäpers[1,2], M. Lyatti[1,2]

[1] Peter Grünberg Institute (PGI-9), Forschungszentrum Jülich, 52425 Jülich, Germany
[2] Jülich Aachen Research Alliance, Fundamentals of Future MaterialsTechnology, 52425 Jülich, Germany
[3] Peter Grünberg Institute (PGI-7), Forschungszentrum Jülich, 52425 Jülich, Germany

* m.lyatti@fz-juelich.de

**Abstract**

Integration of high-temperature superconducting devices with photonic circuits requires a material platform that preserves superconducting properties, provides interfacial heat transport control and optical accessibility. A $YBa_2Cu_3O_{7-x}/SrTiO_3$ (YBCO/STO) nanomembrane platform, enabling integration with $SiO_2$/Si substrates, is developed and characterized. Millimeter-scale ultra-thin YBCO films are grown on a STO/$Sr_{1.5}Ca_{1.5}Al_2O_6$ bilayer, released on a STO nanomembrane, and transferred onto $SiO_2$/Si substrates. X-ray diffraction confirms that crystallinity of STO and YBCO is preserved after transfer. Microbridges patterned from transferred YBCO films exhibit a critical temperature of 88.8 K and a critical current density of 6.8 MA/cm² at 77 K, comparable with those of YBCO films on STO substrates. Statistical measurements show a device fabrication yield of 93%, a critical-temperature variation of only 0.05 K, and 10% variation in critical current density, demonstrating excellent reproducibility. The thermal boundary conductance across the van der Waals interface between the YBCO/STO platform and the $SiO_2$/Si substrate is strongly reduced compared to that of epitaxial YBCO film on STO substrate. The reduced thermal coupling prolongs nonequilibrium states in superconducting nanostructures, which is advantageous for detector applications. These results establish ultra-thin YBCO films on optically transparent STO nanomembranes as a platform for integrating high-$T_c$ superconducting devices with silicon photonics.

## 1. Introduction

Superconducting nanowire single-photon detectors (SNSPD), based on low-temperature superconductors are an established technology. Their performance in the telecommunications frequency range surpasses that of other types of single-photon detectors [1-3]. However, their widespread use in telecommunications is currently limited by the requirement to operate at very low temperatures, in the range of a few degrees Kelvin. The need for such operating temperatures has prompted a search for an alternative superconducting platform that can raise the operating temperature of SNSPDs, thus removing the obstacle to their widespread industrial application.

Recently, SNSPD detectors made from high-temperature (high-$T_c$) superconductors with high operating temperature have been demonstrated [4-6]. The highest operating temperature of up to 25 K has been achieved with $Bi_2Sr_2CaCu_2O_{8+\delta}$-based (BSCCO) detectors making them highly promising [5]. However, BSCCO-based SNSPD fabrication employs mechanical exfoliation together with focused helium ion beam lithography, which poses challenges for scalability and integration with standard device fabrication approaches.

$YBa_2Cu_3O_{7-x}$ (YBCO) is another potential candidate for high-temperature SNSPD. It satisfies key requirements for the single-photon detection. Its critical temperature and critical current density are similar to those of BSCCO [7,8]. Large-area, ultra-thin YBCO films can be grown epitaxially and patterned into nanowires using various lithography techniques [7-12]. The nodal gap of the ultra-thin YBCO can be tuned by varying the film thickness [12]. *IV* curves of YBCO nanowires exhibit the direct voltage switching and current hysteresis required for single-photon detection [7,13,14]. However, despite a few attempts, the YBCO SNSPD has yet to be realized, for reasons that remain unclear.

The works of Charaev *et al.* [5] and Merino *et al.* [4] have highlighted the importance of thermal management for achieving a single-photon response in a cuprate nanowire. In BSCCO-based detectors, thermal coupling to the substrate is reduced by weak van der Waals bonding between the exfoliated flake and the underlying substrate. In $La_{1.55}Sr_{0.45}CuO_4/La_2CuO_4$ devices, thermal transport is suppressed through high-flux helium ion irradiation, which introduces subsurface disorder and modifies heat dissipation pathways. Conversely, high-quality epitaxial $YBa_2Cu_3O_{7-x}$ (YBCO) films typically exhibit relatively high thermal boundary conductance (TBC) at interface with $SrTiO_3$ (STO) substrates [7,15], which may lead to rapid relaxation of nonequilibrium states following photon absorption. An efficient heat removal could suppress the development and persistence of detectable resistive states.

Based on this hypothesis, the realization of YBCO SNSPDs likely requires a material platform with optimized thermal properties and photonic compatibility, for example, free-standing structures or nanomembrane-supported structures in which reduced heat transfer enhances the stability of transient resistive regions. In this context, a platform based on an YBCO film on an optically transparent nanomembrane may be preferable to a fully free-standing YBCO film. Such a nanomembrane protects the ultra-thin YBCO film from degradation while also enabling straightforward integration with photonic circuits. This strategy is consistent with recent advances in hybrid integrated photonic circuits [16].

While several studies have demonstrated free-standing or transferred YBCO films [17-21], these efforts have primarily focused on the preserving the structural properties and the critical temperature $T_c$ after release. This is largely because maintaining in-plane transport in transferred oxide membranes remains challenging task [22]. Thus, application key parameters—such as critical current density, reproducibility of the in-plane superconducting transport, interfacial thermal transport, and compatibility with integration into photonic circuits—remain underexplored.

In this work, we address these operational aspects by developing YBCO/STO nanomembrane platform enabling the integration of YBCO superconducting devices, such as detectors, resonators, filters, and transmission lines, with other devices and platforms. We use an $Sr_{1.5}Ca_{1.5}Al_2O_6$ (SCAO) sacrificial layer to release the YBCO/STO nanomembrane platform and subsequently transfer it onto a Si substrate with a thermally grown $SiO_2$ layer ($SiO_2$/Si substrate). Here, we use the $SiO_2$/Si substrate because $SiO_2$/Si is a standard platform widely used in integrated photonics, enabling straightforward compatibility of YBCO/STO platform with established silicon-based photonic architectures. The high critical temperature and large critical current density together with the X-ray analysis prove the internal in-plane integrity of the transferred YBCO films. We characterize the thermal boundary conductance of the YBCO/STO platform on the $SiO_2$/Si substrate in the 15–75 K temperature range and find that it is reduced by nearly two orders of magnitude compared to epitaxial YBCO on STO substrate, indicating strongly suppressed interfacial heat transport in the nanomembrane geometry. These results establish YBCO films on the STO nanomembrane as a platform with engineered thermal coupling, enabling integration of high-$T_c$ superconducting devices with photonic circuits.

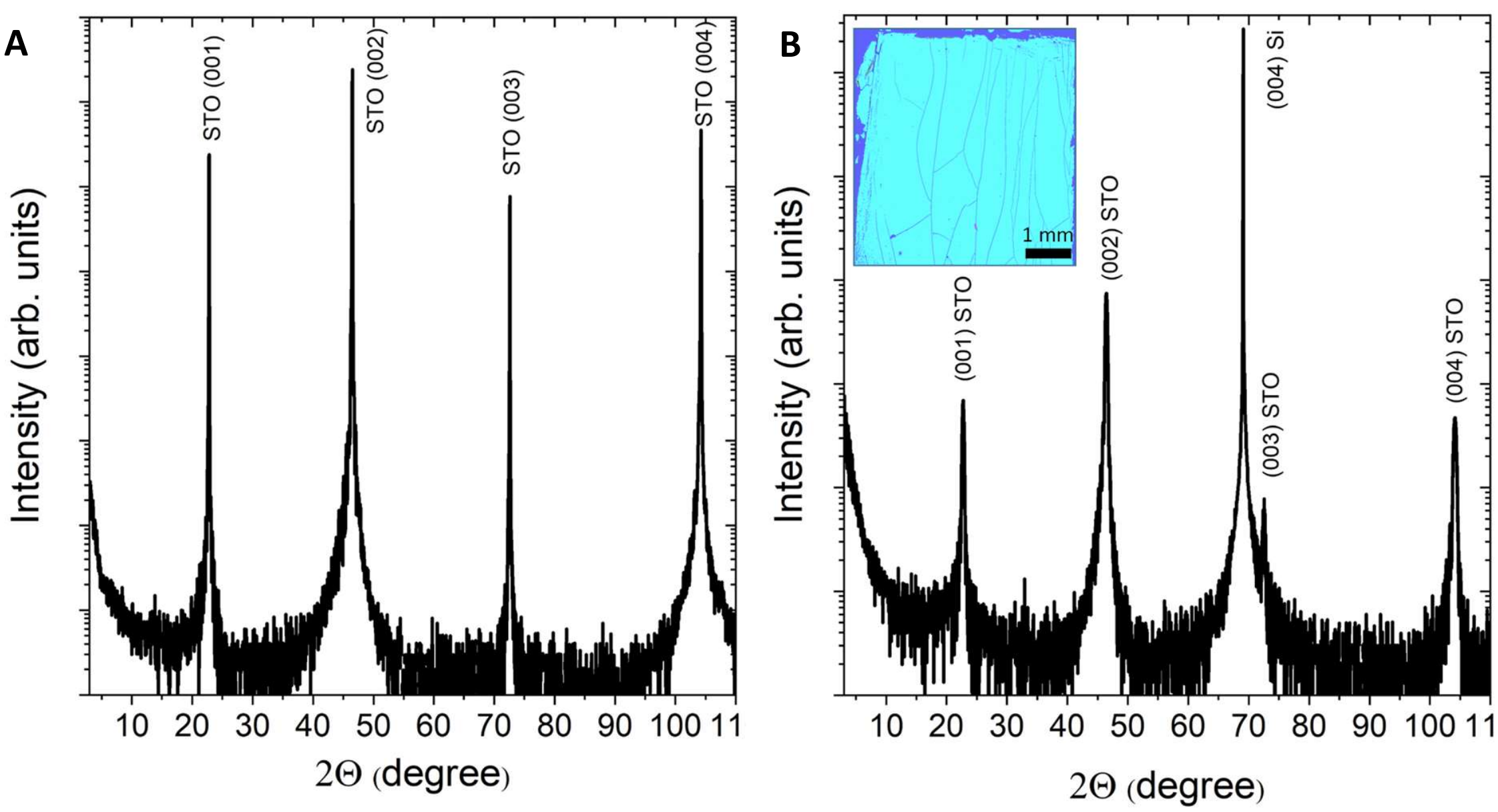

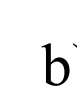

**Figure 1**

A) XRD 2Θ-Θ scans of the STO/SCAO stack on the STO substrate and B) XRD 2Θ-Θ scans of the STO nanomembrane transferred onto the $SiO_2$/Si substrate. Inset in panel B shows an optical image of the STO nanomembrane on the $SiO_2$/Si substrate.

## 2. Results and Discussion

### 2.1. Structural analysis of STO nanomembrane

The first step towards YBCO/STO platform is to fabricate the STO nanomembrane itself and analyze its structural properties. We then deposit the STO/SCAO bilayer on the STO substrate and investigate its crystallographic properties by X-ray diffraction (XRD) analysis using a high-resolution Rigaku SmartLab diffractometer. A reference 2Θ-Θ scan of the STO/SCAO stack is shown in Figure 1a. Only the (00I) reflections of the STO and SCAO are observed in the 2Θ-Θ scan, indicating an epitaxial growth. Reflections of the STO substrate overlap with those of the STO and SCAO epitaxial layers, indicating a good match in lattice constants. The beating pattern of the Laue oscillations around the (00I) peaks (see Figure S1a in Supporting Information) provides evidence of the good crystallinity of the STO and SCAO layers. The STO surface exhibits a step-and-terrace structure with a step height corresponding to half the STO unit-cell height (see Figure S2 in Supporting Information). The matching lattice constants of the STO and SCAO layers and the STO substrate make it difficult to analyze the quality of the STO layer.

Therefore, to measure the lattice constant of the STO layer, we transfer it onto the $SiO_2$/Si substrate using DGL film (Gel-Pak) [23], as described in the "Methods" section, and then perform XRD analysis. The 2Θ-Θ scan of the STO nanomembrane on the $SiO_2$/Si substrate is shown in Figure 1b. A series of STO peaks ranging from (001) to (004), accompanied by Laue oscillations (see Figure S1b in Supporting Information), are observed. The period of the Laue oscillations corresponds to the film thickness determined by X-ray reflectivity (XRR) measurements, and their presence confirms good crystallinity throughout the film. The lattice constant of the STO nanomembrane is equal to 3.906 Å, which is very close to the STO substrate lattice constant of 3.905 Å. This provides evidence of a low defect density in the STO nanomembrane. The STO membrane exhibits cracks following transfer, as illustrated in the inset of Figure 1b. As the spacing between the cracks is strongly anisotropic, it is likely that the cracks are caused by residual stress in the DGL film — stress that remained in the polymer film after it was peeled off from the backing layer and did not completely relax within 30 minutes (see section "Methods"). Nevertheless, the space between these cracks is large enough to pattern microbridges with the current leads.

**2.2. Structural analysis of YBCO film**

Next, we deposit YBCO and capping Au layers atop the STO layer. Figure 2a shows the 2Θ-Θ scan of the Au/YBCO/STO/SCAO stack on the STO substrate. A series of YBCO peaks ranging from (001) to (0011), accompanied by Laue oscillations, are observed. The rocking curve measured for the YBCO film exhibit a full width at half maximum of 0.014° for the (001) reflection, approaching the values reported for high-quality YBCO single crystals [24]. In addition to the (00I) STO, SCAO, and YBCO reflections, the (111) and (222) reflections of the gold are also present, providing evidence of epitaxial gold growth on YBCO, which is in agreement with our previous findings [25]. The c-axis lattice constant of the 20-nm-thick YBCO layer deposited on the STO layer is 11.661±0.007 Å, as determined from the positions of the YBCO XRD peaks. The period of the Laue oscillations around the YBCO peaks (see Figure S3 in Supporting Information) corresponds to the film thickness determined from the deposition time, confirming good crystallinity throughout the film. The density of precipitates on YBCO films deposited on the STO/SCAO bilayer structure is only $3\cdot10^{-3}$ $\mu m^2$, which allows for the fabrication of nanowire meander structures of a size suitable for the SNSPD applications. The 5x5 mm Au/YBCO/STO stack was transferred onto the $SiO_2$/Si substrate using the procedure described in the "Methods" section. The 2Θ-Θ scan of the Au/YBCO/STO stack on the $SiO_2$/Si substrate is shown in Figure 2b. The difference in signal-to-noise ratio between Figures 2a and 2b is due to the use of different detector filters for the 2Θ-Θ scans. A series of YBCO peaks ranging from (001) to (0011) can be clearly identified. The c-axis lattice constant of the YBCO layer on the STO membrane transferred onto the $SiO_2$/Si substrate is

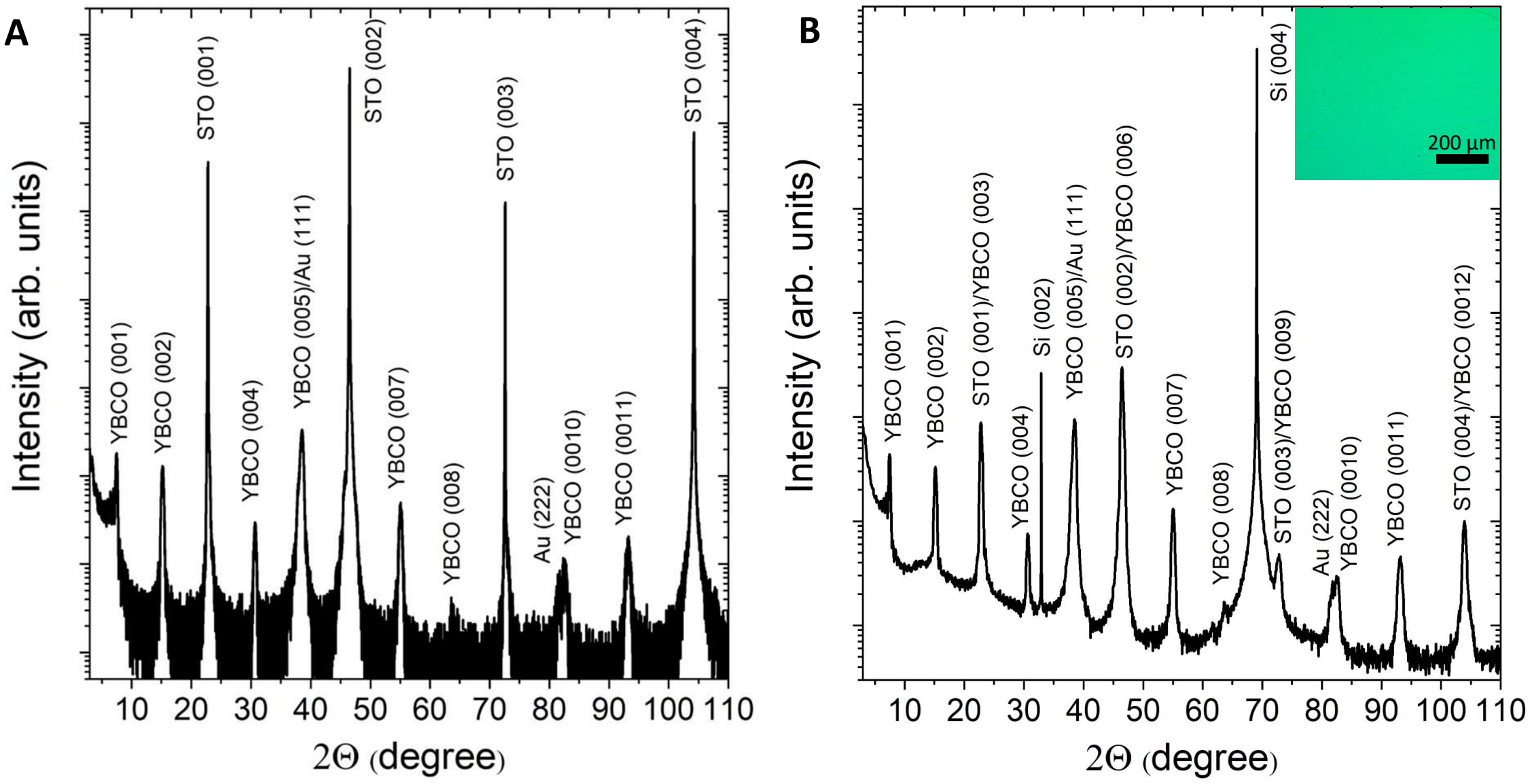

**Figure 2**

A) XRD 2Θ-Θ scans of the Au/YBCO/STO/SCAO stack on the STO substrate and B) XRD 2Θ-Θ scans of the Au/YBCO/STO stack transferred onto the $SiO_2$/Si substrate (b). Inset in panel B shows an optical image of the Au/YBCO/STO heterostructure transferred onto the $SiO_2$/Si substrate.

11.668±0.008 Å, suggesting at most a minor relaxation towards the table value of 11.68 Å for the optimally-doped YBCO. X-ray measurements confirm the high quality of the STO and YBCO layers fabricated by high-pressure sputtering and shows nearly no changes in the film properties of these layers after transfer. The Au/YBCO/STO heterostructure has a smaller spacing between cracks than the STO nanomembrane, as can be seen in the inset in Figure 2b (for a larger scale image, see Figure S4 in Supporting Information). However, the space between these cracks is still large enough to accommodate the microbridges with current leads.

### 2.3. Electrical transport measurements

In the final part of our study, we characterize the superconducting properties of the YBCO/STO platform transferred onto $SiO_2$/Si substrates by patterning the YBCO film into seven µm-sized bridges with 60 µm distance between them and measuring their electrical transport properties. Figure 3 shows an optical image of one of the bridges. The electrical parameters of the Au/YBCO bridges are measured using a 4-probe technique, as illustrated in Figure 3.

Figure 4a shows the typical temperature dependence of the resistance of an Au/YBCO microbridge transferred onto $SiO_2$/Si substrate. The microbridge demonstrate a high critical temperature of 88.8 K (50% resistance drop criterion). This critical temperature value is the same as the one we measure for YBCO films of the same thickness on STO substrates [9]. The

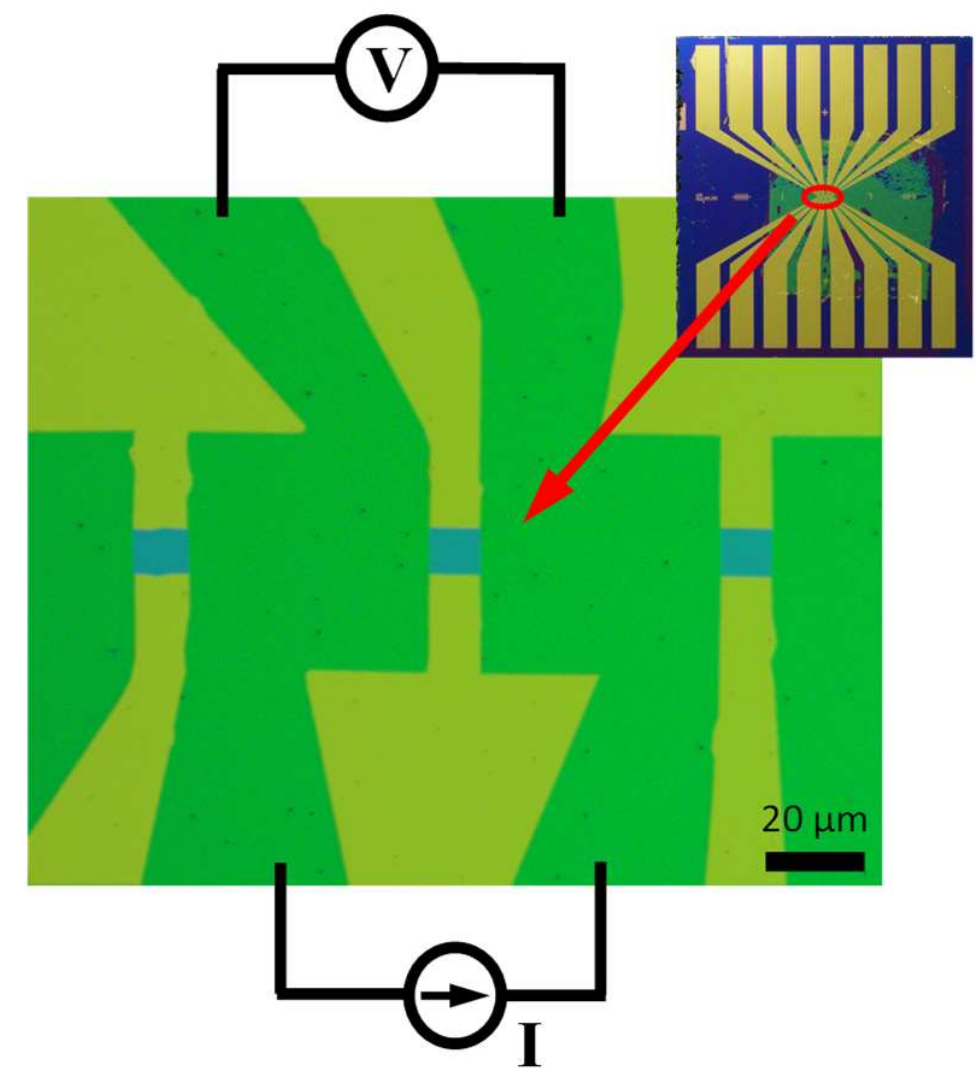

**Figure 3**

A false-color optical image of the Au/YBCO microbridges on the STO nanomembrane transferred onto the $SiO_2$/Si substrate. The Au/YBCO film, the Au/YBCO microbridges, and the STO substrate are highlighted in green, blue, and yellow, respectively. Black dots are precipitates. The inset shows an optical image of a 10x10 mm $SiO_2$/Si chip with seven Au/YBCO microbridges on the STO nanomembrane, as well as gold current leads.

superconducting transition width is 0.5 K (90%–10% resistance drop criterion).The spread of the critical temperatures of the microbridges on the same chip is below 0.05 K (see Figure S5 in Supporting Information) comparable to that observed in Tb-doped YBCO films on STO substrates [26].

Figure 4b shows the typical *IV* curve of an 8.7-μm-wide Au/YBCO bridge on the $SiO_2$/Si substrate with critical current $I_c$ of 8.6 mA. The critical current is determined using the 10 μV criterion. The critical current density of the bridges with the thickness $d$ is calculated as $J_c = I_c/W{\cdot}d$, where $W$ and $d$ are the width and the thickness of the microbridge, respectively. The average critical current density of the 20-nm-thick Au/YBCO microbridges at $T$ = 77.5 K is 4.5±0.4 $MA/cm^2$ for the YBCO film sputtered on the STO layer with a mixed SrO/$TiO_2$ surface termination obtained after STO layer deposition (Sample 1). When the surface of the STO layer is etched using buffered oxide etch (BOE) to obtain a $TiO_2$ surface termination before the YBCO layer deposition, the critical current density increases to 6.8±0.7 $MA/cm^2$ at 77.8 K (Sample 2). These values are approximately one order of magnitude higher than the critical current densities earlier reported for transferred $GdBa_2Cu_3O_{7-x}$ films measured at the significantly lower temperature of 30 K [27] and similar to those measured for YBCO epitaxial films on STO substrates using microbridges of comparable width [28,29]. Additional data are provided in the Supporting Information (Figures S6 and S7). Remarkably, superconducting transport properties are successfully measured in 13 out of 14 fabricated microbridges, corresponding to a device yield of 93%.

The high critical current density together with small spreads in the critical current density and the critical temperature demonstrate that the transfer process preserves in-plane

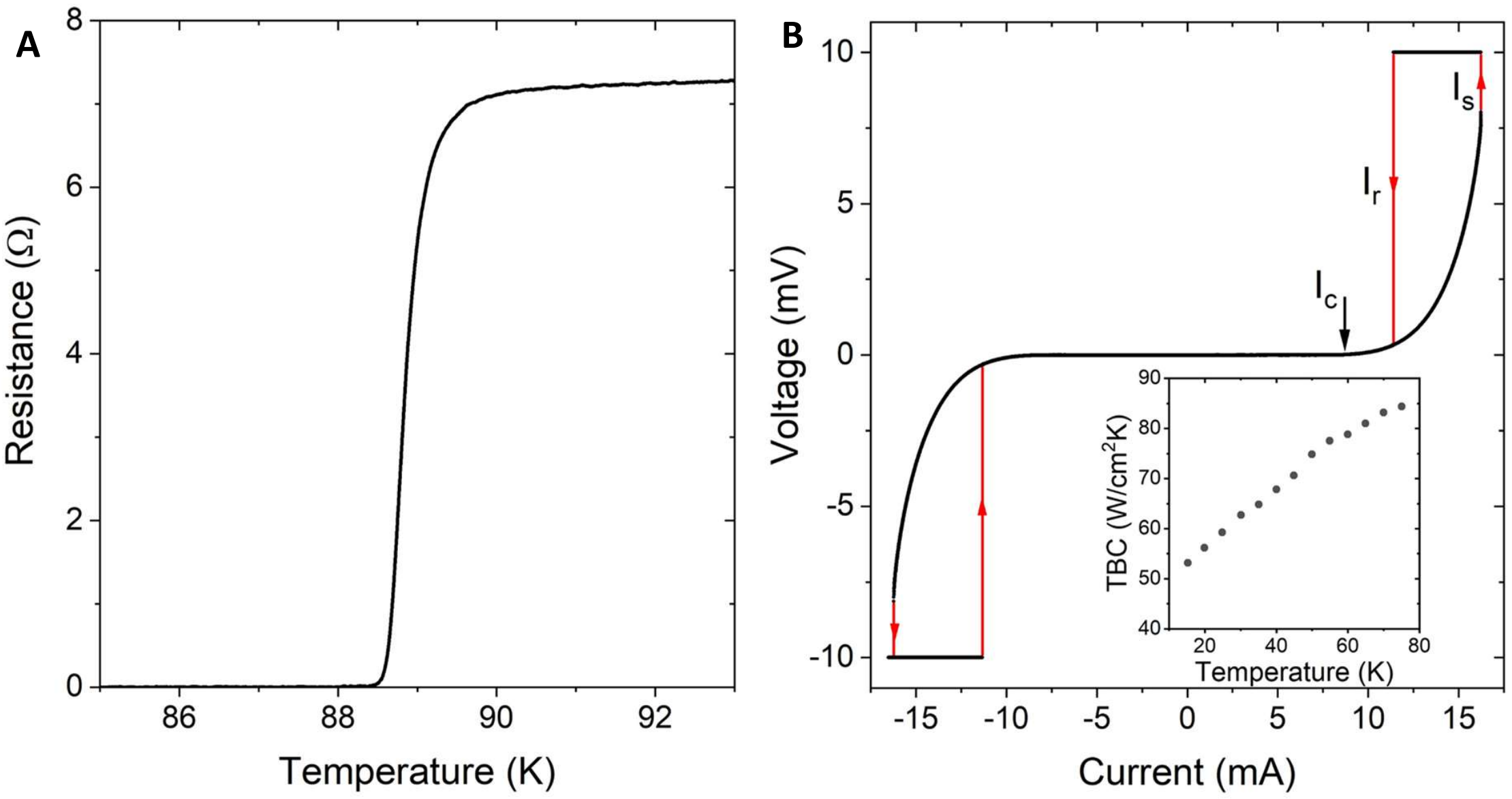

**Figure 4**

A) Resistive superconducting transition of the Au/YBCO bridge on a $SiO_2$/Si substrate. B) *IV* curve at $T = 77.5$ K. The red lines in the inset indicate the voltage switching.

superconducting properties of the ultra-thin YBCO films. These results confirm the structural and electrical transport integrity of the YBCO/STO nanomembrane platform and establish its suitability for the integration of high-performance superconducting devices with $SiO_2$/Si-based photonic circuits.

The *IV* curves of the Au/YBCO bridges on the $SiO_2$/Si substrate demonstrate pronounced switching behavior with current hysteresis, which suggests reduced heat transfer between the YBCO film and the substrate. The abrupt transition to the high-voltage state occurs when the bias current $I_b$ exceeds the switching current $I_s$, which is associated with a flux-flow instability [30]. The appearance of hysteresis indicates that this instability is accompanied by self-heating and the formation of a stable normal-state region [30]. Upon decreasing the current, the system remains in the resistive state until the bias current falls below the retrapping current $I_r$, at which point the bridge cools down below $T_c$ and returns to the low-voltage state.

As discussed in the Introduction section, the heat transfer between the YBCO/STO platform and substrate can greatly affect the detector performance. However, the data on thermal boundary conductance between STO nanomembrane and $SiO_2$/Si substrate at cryogenic temperatures are not available in literature. Therefore, we estimate its values from measurements of the retrapping current. For this purpose, we employ a simplified thermal model that assumes that the heat removal is dominated by heat transfer from the YBCO film into the substrate, while heat flow along the film into the electrodes is neglected. Then, the heat balance at $I_b = I_r$ can be written as $I_r^2 \cdot R_n = \alpha W L \Delta T$ (1), where $\alpha$ is the thermal boundary conductance, the $R_n$ is the bridge resistance at the onset of the superconducting transition, $I_b$ is the bias current, $\Delta T = T_c - T$ is the temperature difference, $T$ is the sample temperature, and $L$ and $W$ are the length and width of the bridge, respectively [31]. Using equation (1), we estimate the TBC between the STO nanomembrane with the YBCO film and the $SiO_2$/Si substrate to be $\alpha \approx 85$ W/cm$^2$K at $T = 77.5$ K. The extracted TBC decreases gradually with decreasing temperature, as shown in the inset in Figure 4b. While this model provides an approximate evaluation, the obtained values are consistent with values reported for van der Waals heterostructures such as $MoS_2/SiO_2$ [32], and is about two orders of magnitude lower than that of epitaxial YBCO films on STO substrates [7,15]. Additional temperature-dependent data are provided in the Supporting Information (Figure S8).

Finally, to illustrate the implications of the engineered thermal coupling provided by the YBCO/STO platform, we analyze its effect on the dynamics of YBCO nanowires, a device geometry widely employed in single-photon detectors for integrated photonic circuits. Within the models describing the SNSPDs, photon absorption generates nonequilibrium quasiparticles which locally suppress the superconducting order parameter [1]. This perturbation trigger vortex entry or hotspot appearance, leading to the formation of a transient resistive domain in the nanowire. The subsequent evolution and stability of this resistive domain are governed by energy relaxation processes, including electron-electron and electron-phonon interaction. Then, in high-$T_c$ superconductors, the TBC may play a critical role in determining whether the transient resistive domain rapidly relaxes back to superconducting state or evolves into a detectable resistive signal. The reduction in thermal conductivity between the high-$T_c$ nanowire with and the substrate enhances energy confinement within the nanowire, promoting the expansion of resistive regions and enabling the detection of single-photon responses.

Analyzing the influence of the reduced TBC on the dynamics of YBCO nanowires, we take into account that in ultra-thin YBCO nanowires with a width of less than 100 nm, a direct transition from the superconducting state to the resistive state occurs via phase slippage [7]. During phase slip, the supercurrent is locally converted into a normal current [33], which injects non-equilibrium quasiparticles into the surrounding YBCO film at a distance of $\Lambda_E = 200$ nm from the phase-slip center [7]. Here, $\Lambda_E$ is the electric field penetration depth. We calculate the thickness $d$ of the YBCO film required to raise the local temperature of the nanowire above $T_c$ following the phase-slip line appearance as $d \approx 2\alpha(4\text{K}) \cdot \Lambda_E \cdot (T_c - 4\text{K})/(V_s \cdot J_c) \approx$ 6-7 nm using parameters of the phase-slip nanowire from our previous work Ref. [7]: switching voltage $V_s =$ 3 mV and critical current density $J_c$ =92.5 MA/cm$^2$ at $T$ = 4 K. Here we assume that bias current is close to critical current. This value falls within the range of film thicknesses suitable for single-photon detection.

## 3. Conclusion

We demonstrate YBCO/STO nanomembrane platform enabling integration of YBCO devices such as detectors, resonators, filters, transmission lines with technologically incompatible platforms. We characterize the platform in terms of structural integrity, superconducting in-plane transport properties, device reproducibility, interfacial thermal transport, and integration with $SiO_2$/Si photonic platform. We grow millimeter-scale ultra-thin YBCO films on an STO/SCAO bilayer and transfer onto $SiO_2$/Si substrates which is a standard platform for integrated photonics. X-ray diffraction confirms that the crystallinity of both the STO nanomembrane and the YBCO layer is preserved after transfer. Superconducting microbridges fabricated out of the transferred Au/YBCO/STO stack exhibit a high critical temperature of 88.8 K and large average critical current densities up to 6.8 MA/cm$^2$ at 77 K. Device-to-device measurements show excellent reproducibility across the crack-free area, with a 93% fabrication yield, a narrow spread in critical temperatures of 0.05 K, and 10% variation in critical current density, indicating the in-plane integrity of the YBCO film. The thermal conductivity at the interface between the YBCO/STO film and the $SiO_2$/Si substrate is reduced by nearly two orders of magnitude compared to an epitaxial YBCO film on an STO substrate, indicating a significant suppression of heat transfer between the superconducting film and the substrate due to the van der Waals interface which is favorable for the single-photon detection.

These results establish the YBCO/STO nanomembrane system as a platform with defined superconducting and thermal properties, suitable for integration of high-$T_c$ superconducting devices with silicon-based photonic circuits.

## 4. Methods

### 4.1. Thin film deposition

First, an STO/SCAO stack was deposited on a 10x10 mm (100) STO substrate. The STO substrate was etched using a BOE and annealed in the sputtering machine at a substrate heater temperature of 920 °C and an oxygen pressure of 1 mbar for one hour, in order to obtain a $TiO_2$-terminated surface. We chose SCAO as a water-soluble sacrificial layer with a lattice constant that is identical to that of STO [34]. The SCAO sputtering pressure was chosen according to the stability diagram reported by Zhang *et al.* [35]. The 11-nm-thick SCAO layer was deposited at an oxygen pressure of 0.05 mbar and a substrate heater temperature of 850 °C by rf sputtering.

A 25-nm-thick STO layer was deposited *in situ* atop the SCAO layer at the oxygen pressure of 1.5 mbar and the substrate heater temperature of the 920 °C by rf sputtering. The deposition rate of STO layer was limited to 6.2 nm/h to achieve a film lattice constant close to the STO table value [36]. The STO/SCAO stack was characterized by XRD and atomic force microscopy to verify crystalline quality. The STO substrate with the STO/SCAO stack atop was then cleaned by sonication in acetone and ethanol, after which it was placed in the sputtering machine. At this stage, a 20-nm-thick YBCO film was deposited *ex situ* atop the STO/SCAO stack at the oxygen pressure of 3.4 mbar and the substrate heater temperature of the 920 °C by dc sputtering. Finally, the YBCO layer was capped *in situ* with a 10-nm-thick gold layer to: (i) protect the YBCO film during the lithography steps; (ii) ensure good electrical contact with the YBCO film; and (iii) prevent damage to the YBCO bridge by overheating in the resistive state. The gold layer was deposited by dc sputtering at the substrate heater temperature of 70 °C. The STO substrate with the Au/YBCO/STO/SCAO stack was finally diced into four 5x5 mm pieces for the further release and transfer of the Au/YBCO/STO stack.

### 4.2. Transfer of STO nanomembrane with Au/YBCO stack.

Transferring the ultra-thin YBCO films is a significant step towards developing an YBCO/STO nanomembrane platform. Several techniques exist for transferring the oxide nanomembrane from one substrate to another [37]. In this study, we employed a technique involving a carrier layer made of an adhesive polymer. Oxide nanomembranes are known to crack or wrinkle when transferred using an adhesive polymer film [22]. This cracking or wrinkling of the nanomembrane occurs due to stress in the multilayer nanomembrane caused by differences in thermal expansion coefficients, lattice constants, and crystallographic defects in the nanomembrane, as well as stress in the polymer film [34,38]. Our aim was to achieve an average distance of several hundred micrometers between cracks, to allow the high-$T_c$ superconducting device with contact pads to be patterned in the crack-free area of the nanomembrane.

During our initial attempts to transfer the STO nanomembranes, we observed a dense anisotropic cracking pattern, which led us to conclusion that stress in the adhesive polymer film could not be ignored. Our observations are consistent with the findings of Yoon et al. [34], who found that a reduction in the number of cracks or wrinkles occurs at the minimum polymer thickness as it allows faster stress relaxation. We also noted that nanomembranes transferred using thicker PDMS films exhibit a higher density of cracks or folds due to slower stress relaxation in the thicker film. Therefore, the stress in the polymer film must be relaxed before attaching it to the substrate with the nanomembrane.

In our work we used a commercially available 6.5-mil-thick DGL series film by Gel Pak [23]. Two retention levels, X0 and X4, were tested. X4 films have stronger adhesion than X0 films. However, the X4 films leave a residue that complicates the subsequent chemical etching and fabrication of low-resistance ohmic contacts. Therefore, our focus is on using DGL film with an X0 retention level.

The potassium hydroxide (KOH) solution with *p*H = 12, 2% Hellmanex III solution with pH = 12, and deionized (DI) water were used to dissolve sacrificial SCAO layer. Although KOH and Hellmanex III solutions enabled faster dissolution of the sacrificial layer, nanomembranes released in these solutions exhibited a higher density of cracks. In contrast, membranes released using DI water showed significantly fewer cracks and improved structural integrity.

Based on these observations, we optimized the transfer process by tuning both polymer stress and sacrificial-layer dissolution conditions to minimize nanomembrane damage. DI water was selected as the optimal etchant for nanomembrane release. A key modification in the transfer protocol is the pre-conditioning of the polymer carrier layer to relax internal stress prior the transfer process**.** The steps of the transfer process are shown in Figure 5. In the first step, a piece of the DGL film of the appropriate size is peeled off from the support film and placed on the DI water surface for half an hour. This step reduces the stress in the DGL film caused by stretching the film when peeling it off the support layer and squeezing the edge of the film with tweezers. In the second step, the substrate with the nanomembrane is placed at the bottom of the Petri dish. The DGL film is then brought into contact with the substrate containing the nanomembrane by controlled water-level adjustment. A polycarbonate (PC) support layer is subsequently applied to stabilize the stack during handling. In the third stage, the PC film/ DGL film/substrate stack is turned upside down so that the substrate is facing up. The Petri dish is then filled with DI water and left until the sacrificial layer has completely dissolved at room temperature over 48 hours. Once the sacrificial layer has dissolved, the $SrTiO_3$ substrate is removed with tweezers. The floating PC/DGL film with the Au/YBCO/STO stack is flipped so that the Au/YBCO/STO layer faces down and placed back on the water's surface. A $SiO_2$/Si substrate is placed at the bottom of the Petri dish, after which the water is removed in order to attach the DGL film with the Au/YBCO/STO stack to the $SiO_2$/Si substrate. The $SiO_2$/Si substrate with the DGL film and the Au/YBCO/STO stack atop is then left drying at 35 °C. In the final step, the DGL and PC films are peeled off, leaving the Au/YBCO/STO stack attached to the $SiO_2$/Si substrate.

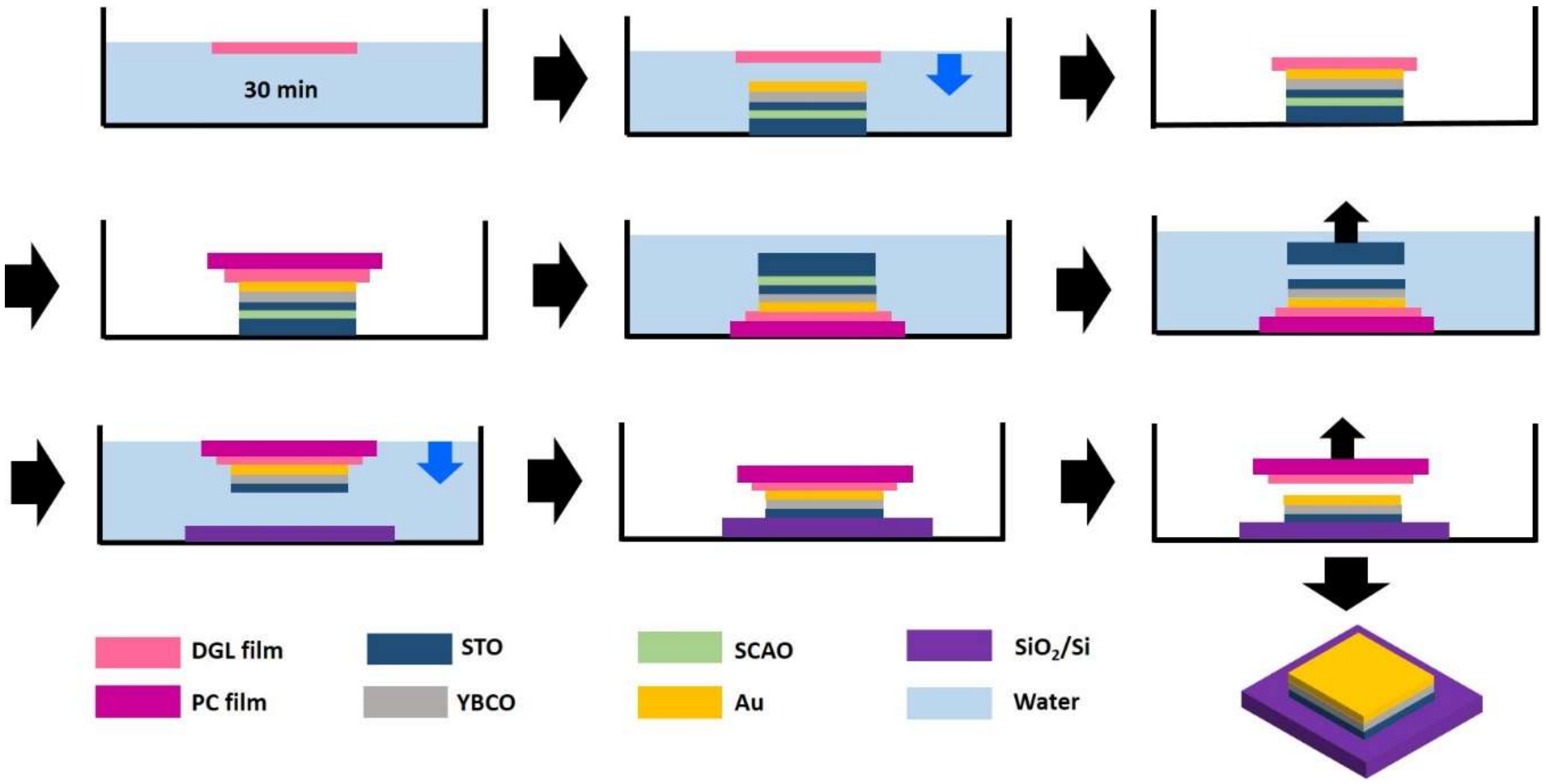

**Figure 5**

Procedure of Au/YBCO/STO stack transfer using DGL film.

### 4.3. Microbridge patterning

The Au/YBCO stack on the $SiO_2$/Si substrate was patterned into seven µm-sized bridges using contact UV contact lithography with PMMA resist, followed by chemical wet etching in (NaI+$I_2$)-ethanol and 0.1% Br-ethanol solutions for gold and YBCO, respectively. The current leads of the microbridges were extended using UV contact lithography, followed by electron beam evaporation of gold and a lift-off process, to meet the requirements of a dipstick designed for 10×10 mm substrates. An optical image of the chip with the YBCO bridges on the $SiO_2$/Si substrate is shown in the inset in Figure 3.

### 4.4. Electrical measurements

We performed the electrical characterization of YBCOinfo microbridges at zero magnetic field using a four-probe technique, as illustrated in Figure 3, using a homemade low-noise, battery-driven electronics and a cryogenic inset in a transport liquid helium Dewar. The temperature dependence of the bridge resistance was measured using a lock-in amplifier with a modulation frequency of 10 kHz and an ac current amplitude of 1 µA. The YBCO microbridges were current-biased during electrical transport measurements.

### Supporting Information

The Supporting Information contains additional information on crystallographic properties of the STO and YBCO films and the reproducibility of the sample parameters.

## Author Contributions

**J.S. Madhira:** Sample fabrication (equal); Measurements (equal); Writing – original draft (equal); Writing – review & editing (equal). **Grigory Potemkin:** Sample fabrication (equal); Writing – review & editing (equal). **B. Kardynal:** Writing – original draft (equal); Writing – review & editing (equal); Project administration (equal), Finding acquisition (equal). **Detlev Grützmacher**: Project administration (equal); Resources; Finding acquisition (equal); Writing – original draft (equal); Writing – review & editing (equal). **Thomas Schäpers:** Project administration (equal); Supervision (equal); Writing – original draft (equal); Writing – review & editing (equal). **Matvey Lyatti:** Conceptualization; Thin-film deposition; Sample fabrication (equal); Measurements (equal); Writing – original draft (equal); Writing – review & editing (equal); Supervision (equal); Project administration (equal), Finding acquisition (equal).

## Acknowledgments

This work was supported by the German Federal Ministry of Education and Research (BMBF) under Grant No. 13N16922 within the *Wissenschaftliche Vorprojekte (WiVoPro)* program on quantum technologies and photonics.

## Data Availability Statement

The data that support the findings of this study are available from the corresponding author upon reasonable request.

Supporting Information

## Development and characterization of $YBa_2Cu_3O_{7-x}/SrTiO_3$ nanomembrane platform for silicon photonics integration

*Jayendra Shankar Madhira, Grigory Potemkin, Beata Kardynal, Detlev Grützmacher, Thomas Schäpers, Matvey Lyatti**

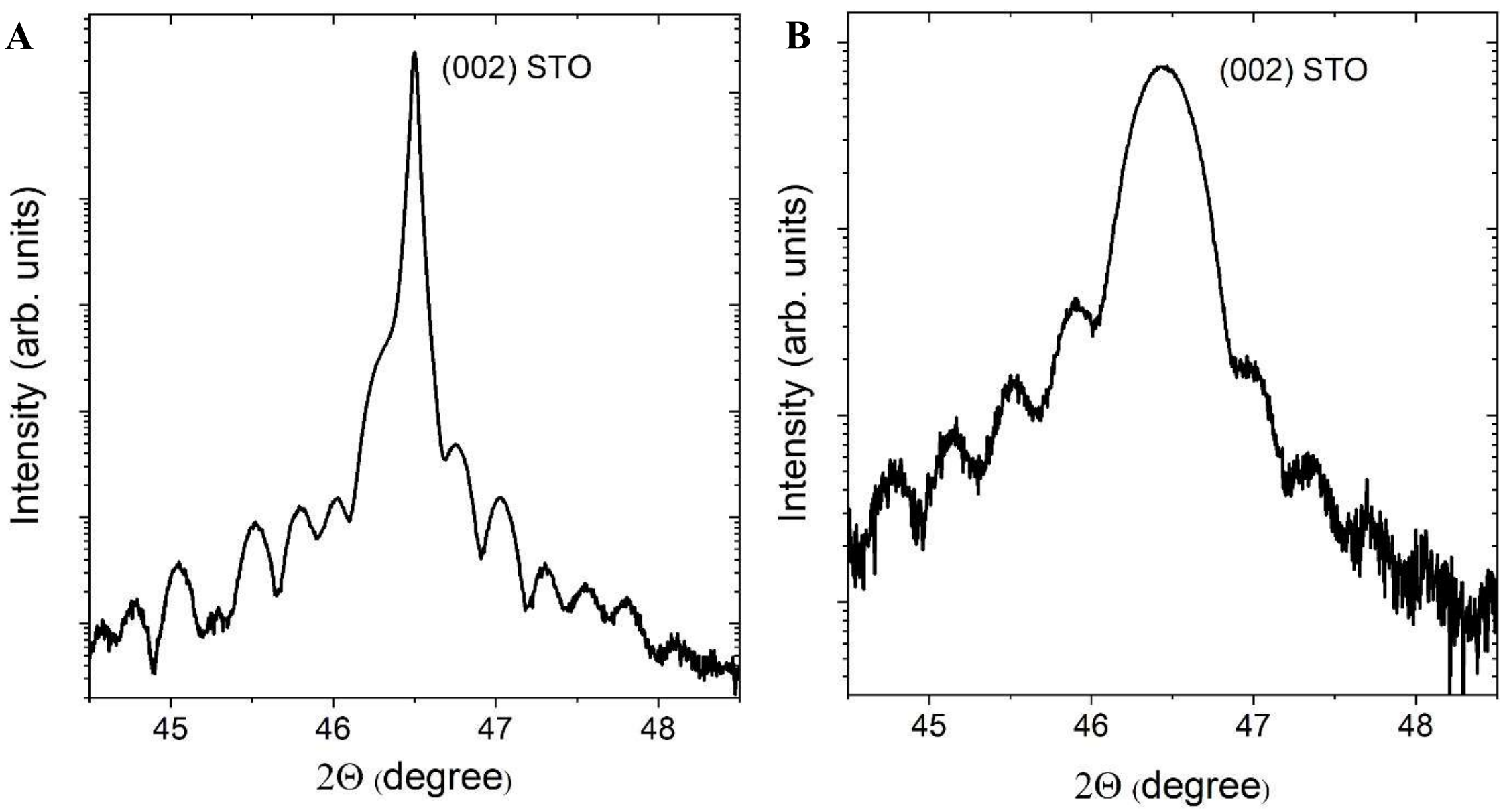

**Figure S1**

A) XRD 2Θ-Θ scan of the $SrTiO_3/Sr_{1.5}Ca_{1.5}Al_2O_6$ bilayer deposited by rf sputtering on the (100) $SrTiO_3$ substrate and B) XRD 2Θ-Θ scan of the $SrTiO_3$ nanomembrane transferred onto the $SiO_2/Si$ substrate.

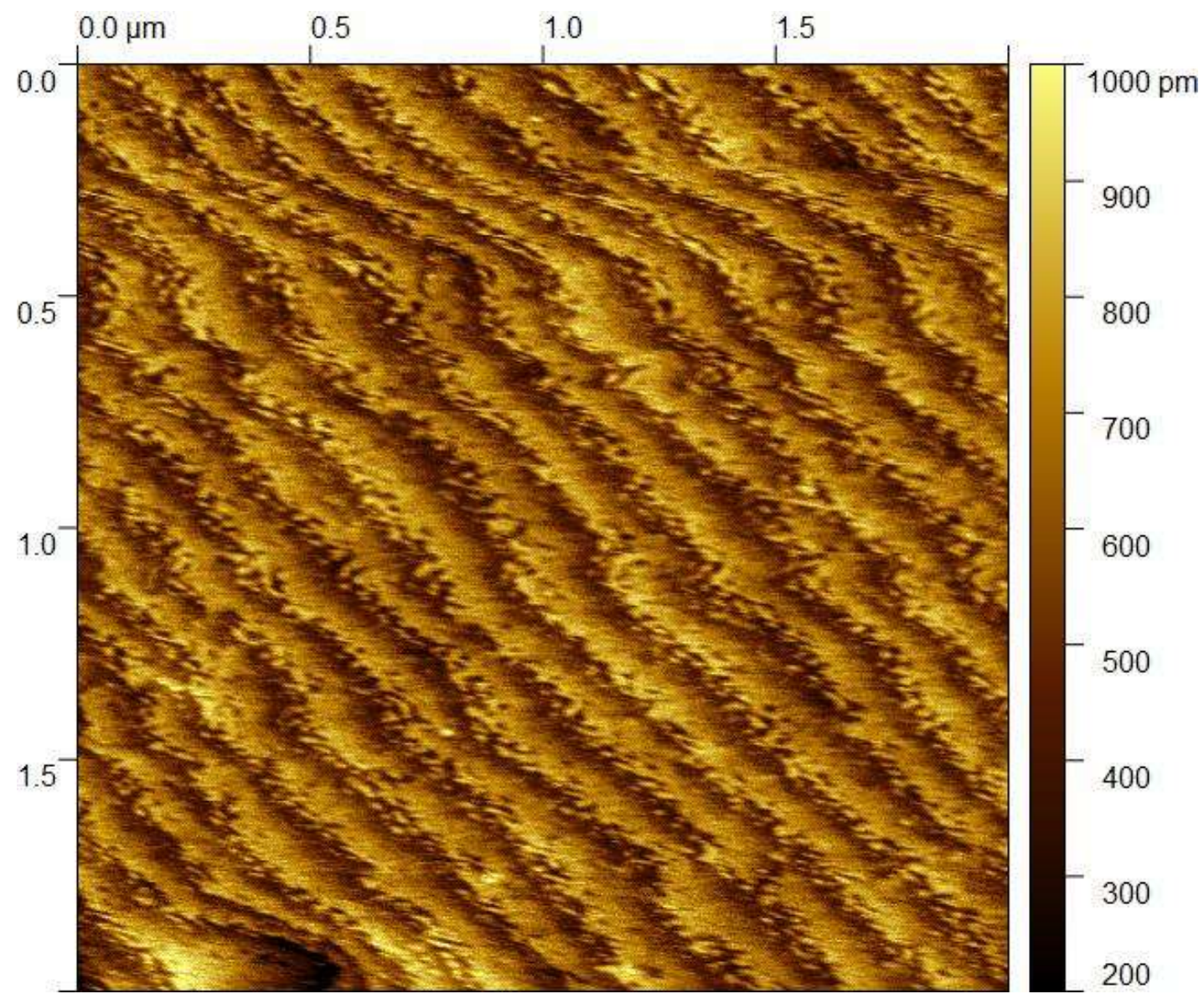

**Figure S2**

The AFM image of the surface $SrTiO_3/Sr_{1.5}Ca_{1.5}Al_2O_6$ bilayer deposited by rf sputtering on the (100) $SrTiO_3$ substrate.

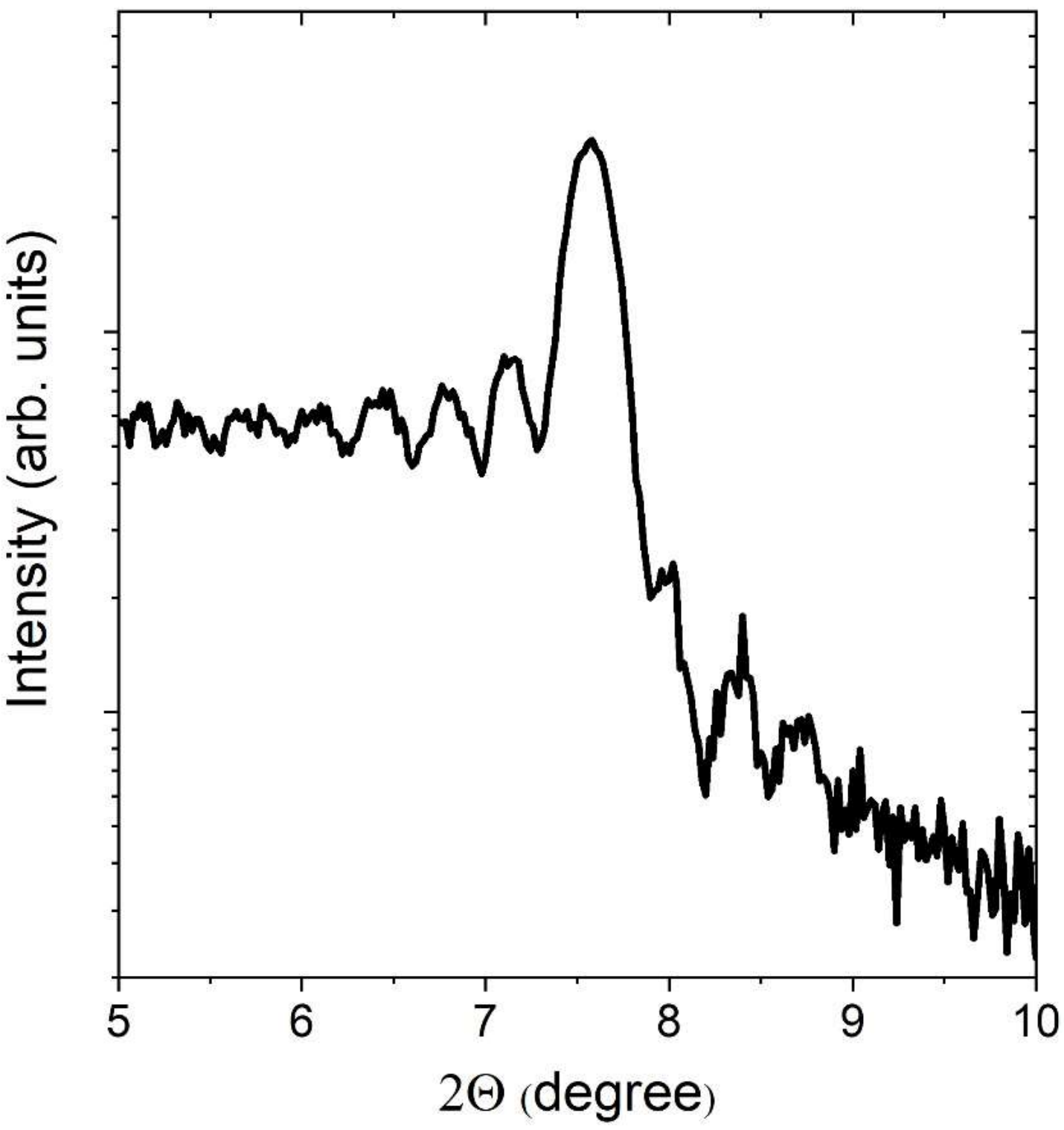

**Figure S3**

XRD 2Θ-Θ scans of the Au/YBCO/STO/SCAO stack on the STO substrate.

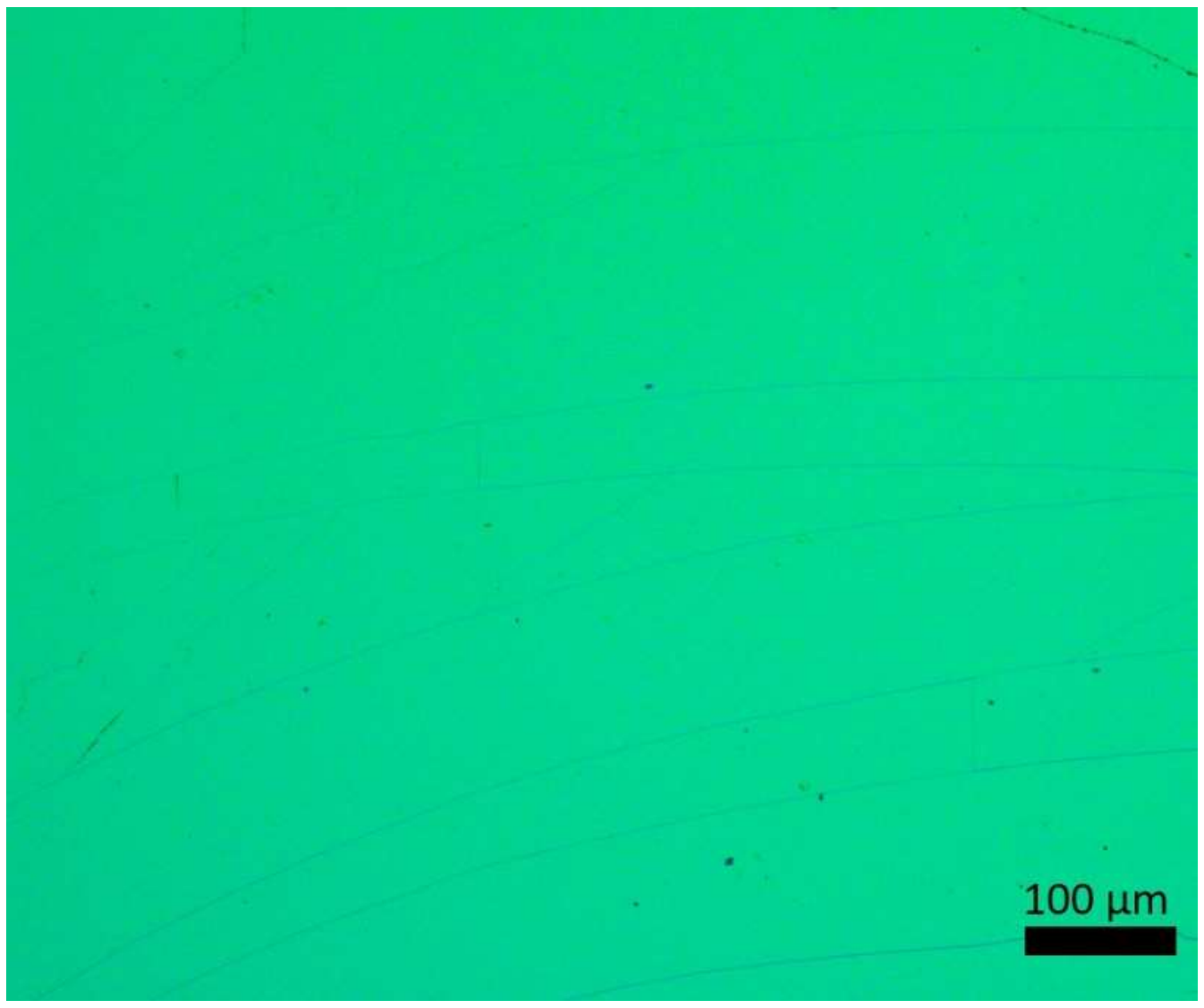

**Figure S4**

An optical image of the Au/YBCO/STO heterostructure transferred on the $SiO_2$/Si substrate.

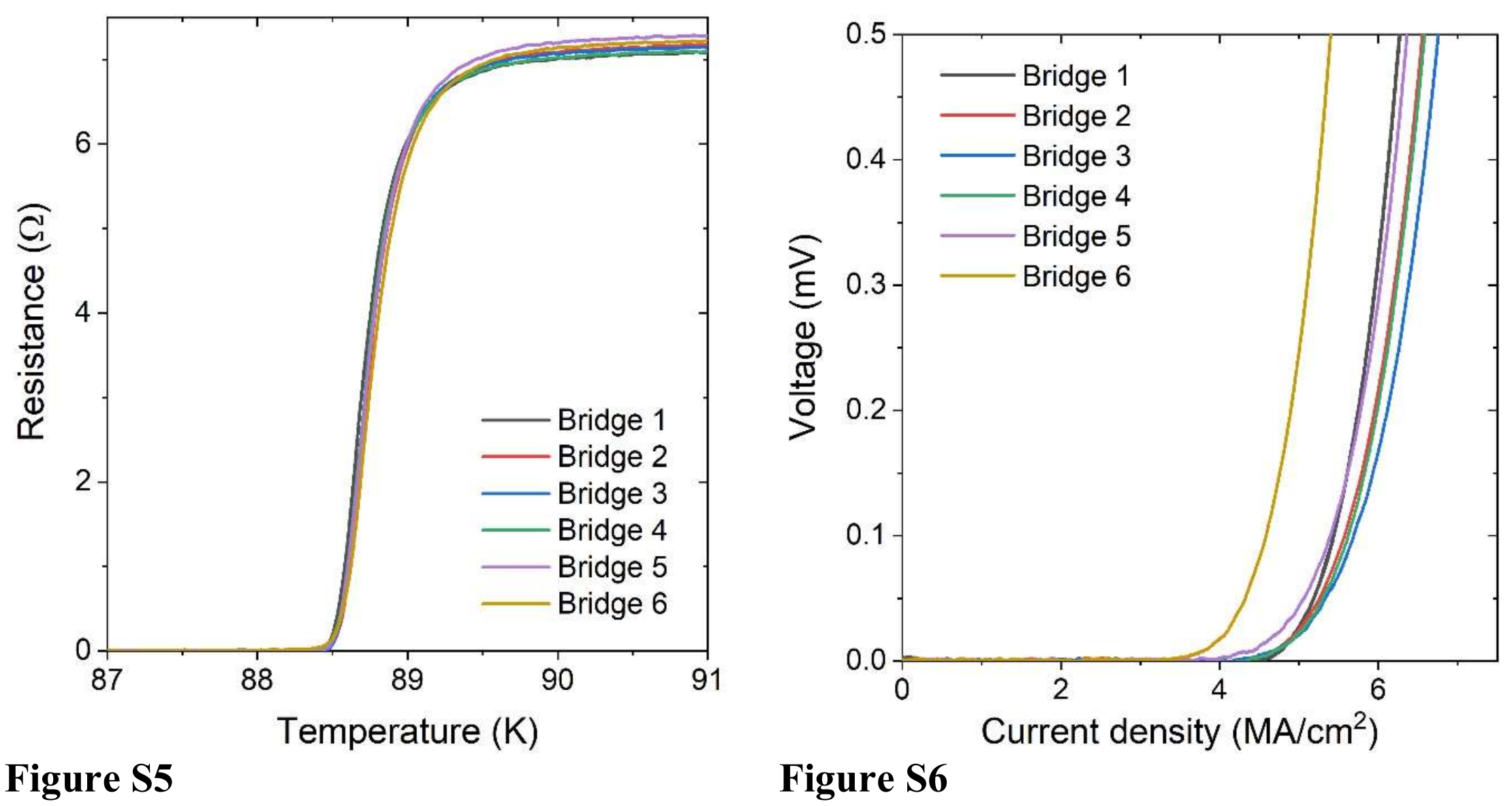

**Figure S5**

Superconducting transitions of the Au/YBCO bridges of the Sample 1 on the $SiO_2$/Si substrate.

**Figure S6**

*IV* curves of the Au/YBCO bridges of the Sample 1 at 77.5 K.

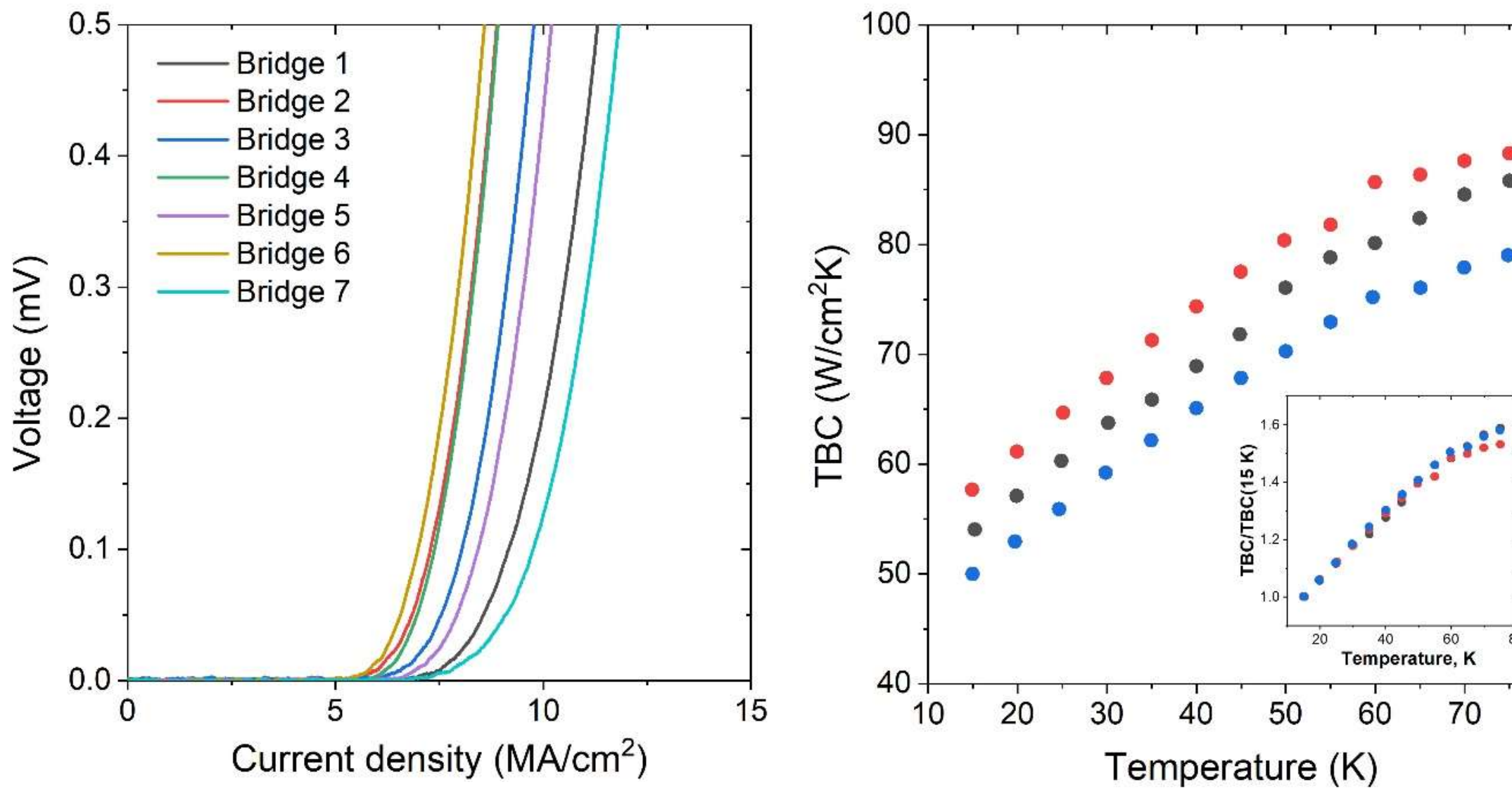

**Figure S7**

*IV* curves of the Au/YBCO bridges the sample 2 on the $SiO_2$/Si substrate at 77.8 K.

**Figure S8**

Temperature dependences of the thermal boundary conductance (TBC) of the Au/YBCO bridge N2 (black dots), N3 (red dots), and N4 (blue dots) of Sample 1. Inset shows the same data normalized to the TBC value at 15 K